\documentclass[12pt,prb,aps,showpacs,preprint]{revtex4}

\usepackage{graphics}
\usepackage{graphicx}
\usepackage{amsfonts}
\usepackage{textcomp}
\usepackage{amssymb}
\usepackage{amsmath}
\usepackage{color}
\usepackage{ulem}

\begin{document}

\title{Non-Universality of Density and Disorder in Jammed Sphere Packings}

\author{Yang Jiao}


\affiliation{Department of Mechanical and Aerospace Engineering,
Princeton University, Princeton New Jersey 08544, USA}

\author{Frank H. Stillinger}


\affiliation{Department of Chemistry, Princeton University,
Princeton New Jersey 08544, USA}

\author{Salvatore Torquato}

\email{torquato@electron.princeton.edu}

\affiliation{Department of Chemistry, Princeton University,
Princeton New Jersey 08544, USA}

\affiliation{Department of Physics, Princeton University,
Princeton New Jersey 08544, USA}

\affiliation{Princeton Center for Theoretical Science, Princeton
University, Princeton New Jersey 08544, USA}

\affiliation{Program in Applied and Computational Mathematics,
Princeton University, Princeton New Jersey 08544, USA}

\date{\today}


\begin{abstract}

We show for the first time that collectively jammed disordered
packings of three-dimensional monodisperse frictionless hard
spheres can be produced and tuned using a novel numerical protocol
with packing density $\phi$ as low as 0.6. This is well below
the value of 0.64 associated with the maximally random jammed
state and entirely unrelated to the ill-defined ``random loose
packing'' state density. Specifically, collectively jammed
packings are generated with a very narrow distribution centered at
any density $\phi$ over a wide density range $\phi \in
[0.6,~0.74048\ldots]$ with variable disorder. Our results support
the view that there is no universal jamming point that is
distinguishable based on the packing density and frequency of
occurence. Our jammed packings are mapped onto a
density-order-metric plane, which provides a broader
characterization of packings than density alone. Other packing
characteristics, such as the pair correlation function, average
contact number and fraction of rattlers are quantified and
discussed.

\end{abstract}

\maketitle

\section{Introduction}

The fundamental study of disordered jammed packings has a long
history dating back at least to the pioneering work of Bernal
\cite{Be60, Be65} on the so-called random-close-packing (RCP)
state. That was traditionally thought to correspond to the densest
``random" packing with a unique packing fraction (or density)
$\phi \approx 0.64$ for frictionless monodisperse
three-dimensional (3D) spheres. That work spawned substantial
research on disordered jammed packings.
\cite{LSpacking,Sp94,Edwards,Liu98,SalMRJ,SalMRJ2,OHern,Do05g2,Za10,SalReview}
About a decade ago, \cite{SalMRJ} it was argued that the RCP state
is ill-defined for various reasons, including the fact that
``randomness" was never quantified and the idea that a ``random"
packing can ever achieve a maximal density is not meaningful
because one can infinitesimally increase the density of the
putative RCP state, which indeed is dependent on the packing
protocol and container boundaries, with imperceptible change in
the amorphous pair correlation function. It was suggested that the
RCP state should be replaced by the \textit{maximally random
jammed} (MRJ) state, \cite{SalMRJ} which is the one that minimizes
a scalar order metric $\psi$ subject to the condition of the
degree of jamming. \cite{SalJam,SalReview} Studies of different
order metrics for 3D frictionless spheres have consistently led to
a minimum  at approximately the same density $\phi \approx 0.64$,
\cite{SalMRJ,Truskett00,SalMRJ2} for collectively and strictly jammed packings in
the $\phi$-$\psi$ plane (i.e., the ``order map''). \cite{SalJam}
This consistency among the different order metrics speaks to the
utility of the order-metric concept, even if a perfect order
metric has not yet been identified. The fact that this jammed
state is epitomized by maximal disorder has been advocated by
other groups. \cite{OHern} Bernal originally studied such jammed
packings to understand the structure of liquids, but it is now
known that 3D MRJ monodisperse sphere packings possess
quasi-long-range  pair correlations. \cite{Do05g2}  This property
is markedly different from typical liquids with short-range
interactions, which possess pair correlations decaying
exponentially fast. Thus, MRJ sphere packings can be regarded to
be prototypical glasses in that they possess the highest degree of
disorder among all jammed packings with diverging elastic moduli.
\cite{SalTunnel}


The definition of the MRJ state implies that the density $\phi$
alone is not sufficient to characterize jammed packings, since any
packing configuration, jammed or not, is a point in the
$\phi$-$\psi$ plane. This two-parameter description is but a very small
subset of the relevant parameters that are necessary to fully characterize a
configuration, but it  nonetheless enables one to draw important
conclusions. \cite{SalReview} The frequency of occurrence
of a particular configuration is irrelevant insofar as the order
map is concerned, i.e., the order map emphasizes a
``geometric-structure'' approach to analyze packings by
characterizing individual configurations, regardless of their
occurrence probability. \cite{SalReview}

One might argue that the maximum of an appropriate ``entropic"
metric (based on the frequency of occurrence of the packings)
would be an ideal way to characterize the randomness of a
packing and therefore the MRJ state. However, as pointed out by
Ref.~\onlinecite{SalMRJ2}, a substantial hurdle to overcome in
implementing such an order metric is the necessity to generate all
possible jammed states in an unbiased fashion using a ``universal" protocol
in the large-system limit, which is an intractable problem. Even
if such a universal protocol could be developed, the
issue of what weights to assign the resulting configurations
remains. Moreover, there are other fundamental problems with
entropic measures. It is well known that the lack of ``frustration'' \cite{SalReview}
in two-dimensional analogs of three-dimensional
computational and experimental protocols that lead to putative MRJ states
result in packings of monodisperse circular disks that are highly crystalline,
forming rather large triangular coordination domains.
Because such highly ordered packings are the most probable outcomes for these typical
protocols, ``entropic measures'' of disorder would identify these as
the most disordered, which is clearly a misleading conclusion.

On the other hand, an appropriate geometric-structure order metric
is capable of identifying a particular configuration not an
ensemble of configurations of considerably lower density. For example, a
jammed vacancy-diluted triangular lattice packing (and its
multidomain variant) or a jammed packing containing many small
crystalline regions and ``grain'' boundaries that is consistent
with our intuitive notions of maximal disorder 
possesses small scalar order metrics. However, typical
packing protocols would almost never generate such disordered disk
configurations because of their inherent implicit bias toward
undiluted crystallization. The readers are referred to
Ref.~\onlinecite{SalReview} for further discussion. Thus we seek
to devise order metrics that can be applied to single jammed
configurations, as prescribed by the geometric-structure point of
view. The geometric-structure approach incorporates not only
maximally dense packings (e.g., Kepler's conjecture) and random
``Bernal" packings, but an infinite class of other significant
jammed states not previously recognized, including ``tunneled"
crystals that are putatively at the jamming threshold with $\phi
\approx 0.49365\ldots$.

Furthermore, the geometric-structure approach naturally
incorporates the algorithmic variability of different packing
protocols that leads to a diversity of density and disorder in
jammed sphere packings. \cite{SalReview} In a typical numerical
packing protocol, either the particle growth or system compression
leads to an increase of $\phi$, \cite{Sp94, LSpacking, To09,
To09b} which makes the particle-pair nonoverlap constraints
consume larger and larger portions of the configuration space.
\cite{SalReview} Further increase of $\phi$ causes the available
configuration space to fracture, generating isolated ``islands''
that each eventually collapse into final jammed states with
distinct density and structure. Presumably, any protocol would
sample the disconnected regions of the configuration space with
fixed probability, leading to well-defined average of any packing
characteristic of interest. \cite{SalReview} Unless chosen to be
highly restrictive, a typical packing protocol applied to a system
with $N$ particles could produce a large number of geometrically
distinguishable packings with some dispersion in their $\phi$ and
$\psi$ values. A narrowing of the distributions of the packing
characteristics with increasing $N$ can be expected due to
operation of a central limit theorem. The particular values to
which the distributions individually converge are
protocol-specific, meaning that these values can be controlled by
choosing appropriate protocols or tuning the parameters of a
protocol. Indeed, jammed packings with a diversity of density and
disorder have been produced via a variety of protocols,
\cite{SalMRJ, Sp94, Truskett00, SalMRJ2, SalTunnel, Cha10, He09}
including $\phi \approx 0.64$ which is empirically the outcome of
a considerable large number of laboratory experiments and
numerical simulations for identical frictionless spheres. Since
the jammed packings with $\phi \approx 0.64$ are apparently the
``most probable states'', significance has been attached to the
so-called unique J (jamming) point, i.e., $\phi \approx 0.64$, which
is suggested to correspond to the onset of collective jamming in
soft-sphere systems. \cite{OHern} However, the wide spectrum of
density values [e.g., $\phi \in (0.63, 0.74048\ldots)$] that has
been achieved clearly suggests that conclusions drawn from any
particular protocol are highly specific rather than general and
thus claims of uniqueness of packing states based on their
frequency of occurrence overlook the wide variability of packing
algorithms and the distribution of configurations that they
generate. We will elaborate on this issue in the Conclusions and
Discussion (Sec. IV).

In this paper, we show explicitly that exploring algorithmic
variability of packing protocols can lead to a diversity of
density and disorder of jammed sphere packings, which is
consistent with the geometric-structure approach. In particular,
we demonstrate that collectively jammed packings can be generated
with a narrow distribution centered at any density $\phi$ over a
wide range $\phi \in [0.6,~0.74048\ldots]$. A novel sequential
linear programming (SLP) packing algorithm \cite{To10} is used to
produce jammed disordered packings with $\phi$ as low as 0.6 for
the first time, i.e., the {\it onset} of disordered jamming occurs
well below the MRJ density of about 0.64. The
Lubachevsky-Stillinger (LS) packing algorithm \cite{LSpacking} is employed
to produce jammed packings with $\phi$ spanning
continuously from that of MRJ state ($\phi \approx 0.64$) all the
way up to the face-centered-cubic (fcc) close-packed density
($\phi = 0.74048\ldots$). \cite{SalMRJ,Truskett00,SalMRJ2}
However, the standard LS algorithm as well as all previously used
numerical protocols do not produce disordered collectively jammed
states with $\phi$ well below 0.64. We control the jamming density
by tuning certain parameters of the packing protocols. The jammed
packings with a diversity of disorder are mapped onto a
density-order-metric plane. Our results strongly support the view
that there is no universal jamming point that is distinguishable
based on the packing density and its frequency of occurence. Other
packing characteristics, such as the pair correlation function
$g_2$, average contact number $Z$,
 and fraction of rattlers $f_r$ are quantified and discussed.

\section{Packing Protocols and Control Parameters}


\subsection{Lubachevsky-Stillinger Algorithm}

 The LS algorithm is an event-driven molecular dynamics in which particles can grow in
size at a certain expansion rate $\gamma = \frac{1}{2}dD/dt$ ($D$
is the diameter of the sphere) in addition to their thermal motion.
\cite{LSpacking} Note that $\gamma$ is a key parameter that
controls the jamming density. \cite{SalMRJ, Do05g2} Sufficiently
small $\gamma$ enables the system almost to be in equilibrium
during the densification, which will finally crystallize in three dimensions into a
fcc packing. For large $\gamma$, the system will quickly fall out
of equilibrium and reaches a jammed state with an amorphous
structure. Intermediate $\gamma$ will result in various degrees of
partial crystallization in the system, which leads to a continuous
range of jamming densities. It is noteworthy that a very small value of 
$\gamma$ should be used toward the jamming limit such that a true
particle contact network can be formed for both crystalline and
disordered packings. \cite{Do04JamTest, SalReview} We use a
modified version of the LS alogorithm \cite{Do05} to generate
jammed packings for $\phi \ge 0.64$.


\subsection{Sequential-Linear-Programming Algorithm}

 A recently devised SLP algorithm \cite{To10} is used here to produce
disordered jammed sphere packings with $\phi<0.64$. This algorithm
solves an optimization problem called
 the adaptive shrinking cell (ASC) scheme: \cite{To09, To09b}
jammed particle packings are generated by maximizing the packing
density subject to interparticle nonoverlapping constraints. The
optimization variables include particle positions as well as shape
and size of the simulation box. Starting from an initial
configuration, a new configuration is obtained by (locally)
maximizing the density via both individual particle motions and
collective motions induced by the deformation/shrinkage of the
simulation box. For spheres, the objective function and
constraints can be linearized for a given packing configuration
and the SLP  method is used to solve the optimization problem. The
final packings produced by the SLP are at least collectively
jammed due to its incorporation of inherent collective particle
motions. Details of this algorithm 
and its applications for generating both disordered and
maximally dense packings in high dimensional Euclidean space 
are given in Ref.~\onlinecite{To10}.


By removing spheres from the fcc packing and its stacking variants
without destroying the rigidity of the contact network, Torquato
and Stillinger \cite{SalTunnel} have constructed strictly jammed
``tunneled crystal'' packings with $\phi$ as low as
$0.49365\ldots$. Similar removal procedures can be applied to MRJ
packings. However the contact network of the remaining packing is
generally not rigid anymore. The removed particles are required to
have a large coordination number and to be mutually separated by
at least a few sphere diameters. Compressing the remaining packing
using the SLP algorithm leads to a re-jammed configuration, with a
minimum degree of structural relaxation (i.e., significant
particle reconfigurations are localized around the cavities).
\cite{To10} The re-jammed configurations generally possess $\phi
< 0.64$ and the removal-compression procedure can be repeated
several times until a lower limit of $\phi$ is reached. The
control parameters of the SLP algorithm include the initial MRJ
configurations and the number of removed particles.

\section{Jammed Packings with Variable Density and Disorder}

\subsection{Histograms for Jammed Packings with $\phi>0.64$}

We employ the LS algorithm to generate a large number of jammed
packings of monodisperse spheres in three dimensions for $\phi \ge 0.64$ with $N =
250, 500, 1000, 2500, 5000$, and $10000$, and a wide range of
initial expansion rate $\gamma \in [10^{-2},~ 10^{-6}]$. Results
are verified to be at least collectively jammed,
\cite{Do04JamTest} using either a linear programming protocol
\cite{Do04JamTest} or by monitoring the instantaneous pressure of
the systems for long time periods; \cite{Do05g2} and rattlers are
included to compute the reported densities. By tuning $\gamma$, the
density at which the systems jam can be controlled. Figure
\ref{fig1}(b), (c) and (d) contrast distributions of $\phi$ for
two distinctly different system sizes ($N = 250$ and $2500$)
converging onto $\phi \approx 0.64, 0.68$ and $0.72$,
respectively. (The case of $\phi \approx 0.60$ shown in Fig.~1(a)
will be discussed separately below.) It is clear that as $N$
increases the $\phi$ distributions narrow for all three mean
density values. We also find that such narrowing becomes even more
significant for $N = 5000$ and $10000$. Specifically,
Fig.~\ref{fig2} shows how the standard deviation $\sigma$ of the
density distribution varies with system size $N$ for the case in
which the mean $\phi \approx 0.66$. Observe that  $\sigma$ is a
monotonically decreasing function of $N$ and $\sigma$
approximately scales $ N^{-1/2}$ for large $N$. A similar
narrowing of the $\phi$ distribution occurs for $\phi \approx
0.60$ shown in Fig.~1(a), and this case will be discussed
separately below. Thus, one can expect that in the
``thermodynamic'' limit (i.e., $N \rightarrow \infty$) the jamming
density will converge to a well-defined value anywhere over the interval $\phi \in
[0.60,0.74048\ldots]$. These results imply that any inclination to
select a specific $\phi$ value as uniquely significant (e.g.,
$\phi \approx 0.64$) is primarily based on inadequate sampling of
the full range of algorithmic richness and diversity that is
available at least in the underlying mathematical theory of
jamming.

\subsection{Histograms for Jammed Packings with $\phi<0.64$}

 We produce the majority of jammed packings with $\phi<0.64$ using
the  SLP algorithm. MRJ packings generated
 via the LS algorithm are used as initial configurations. Each time
approximately $f_s = 0.1\% - 2.5\%$ of the spheres are removed
from the initial packing and the remaining spheres are compressed
to a jammed state using the SLP algorithm. Such a procedure is
repeated $n_r = 5 - 10$ times before a lower limit on $\phi$ is
reached. The fraction $f_s$ of removed spheres is decreased as the
limit is approached. We stress that our packings with $\phi<0.64$
are not so-called {\it random loose packings}, \cite{RLP} which
are not even collectively jammed. \cite{Do04JamTest}
A few packings with $\phi\approx 0.62$ are
generated using the LS algorithm with open simple-cubic lattice
packings as initial configurations. \cite{SalMRJ2}

Figure \ref{fig1}(a) contrasts $\phi$ distributions for two
distinctly different system sizes ($N = 216$ and $2235$)
converging onto $\phi \approx 0.60$. It can be seen that as $N$
increases the $\phi$ distributions narrow. Such narrowing is also
observed for other convergence densities within the interval $[0.6, ~0.64]$. We
note that it is very difficult to produce jammed packings with
$\phi$ significantly lower than 0.6 using the SLP algorithm
(though removing the rattlers in the packing results in a slightly
lower density $\sim 0.595$), while it has been rigorously shown
that strictly jammed sphere packings (e.g., tunneled
crystals and the associated stacking variants) can possess $\phi$
as low as $\frac{\sqrt2 \pi}{9} = 0.49365\ldots$. \cite{SalTunnel}
However, jammed packings that combine the tunneled crystals and
MRJ packings can be constructed. In particular,
stackings of layers of honeycomb-lattice packings of spheres stabilized with triangular-lattice
layers on top and bottom are inserted into the MRJ packings. These ``layered'' packings
are then compressed to jamming using the SLP algorithm. This construction
enables one to obtain jammed packings with variable disorder within the density
range $\phi \in (0.49, 0.64)$.

\subsection{Order Metrics and Other Packing Characteristics for Jammed Packing with $\phi\in[0.49,~0.74]$}

 It has been established that a variety of different useful order metrics
are positively correlated \cite{SalMRJ, Truskett00, SalMRJ2} and
hence any one of them can be used to characterize the packings. To
quantify the order of the packings, we compute the translational
order metric $\mathfrak T$, \cite{SalMRJ} defined as
\begin{equation}
\label{eq1}
\mathfrak{T} =
\left|{\sum_i^{N_s}(n_i-n_i^{ideal})/\sum_i^{N_s}(n_i^{FCC}-n_i^{ideal})}\right|,
\end{equation}
where $n_i$ is the average occupation number for the shell $i$
centered at a distance from a reference sphere that equals the
$i$th nearest-neighbor separation for the open FCC lattice at that
density and $N_s$ is the total number of shells for the summation
($N_s = 45$ is used here for $N\sim 2500$); $n_i^{ideal}$ and
$n_i^{FCC}$ are the corresponding shell occupation numbers for an
ideal gas (spatially uncorrelated spheres) and the open FCC
lattice. For a completely disordered system (e.g., a Poisson
distribution of points) $\mathfrak{T} = 0$, whereas $\mathfrak{T}
= 1$ for the FCC lattice.

Figure \ref{fig3} shows the $\phi$-$\mathfrak{T}$ plane on which
representative jammed packings with $N\sim2500$ and $\phi\in
[0.49,~0.74]$ are mapped. The packings
generated using the LS algorithm are shown as black circles. Note
that for these packings each $\phi$ is associated with a range of
$\mathfrak{T}$ values. Moreover, increasing $\phi$ from the MRJ
state can be achieved at the cost of decreasing the degree of
disorder, as pointed out in Ref.~\onlinecite{SalMRJ}. Further
increase of $\phi$ is associated with partial crystallization of
the packings, as indicated by the sharp peaks of the pair
correlation function $g_2$ of the packings (see Fig. \ref{fig4}).
The fraction of ``rattlers'' (i.e., locally unjammed individual
particles that can move freely within cages of jammed neighbors)
decreases (from $\sim 2.8\%$ with $\phi \approx 0.64$ to $0\%$
with $\phi = 0.74048\ldots$) and the average contact number per
particle $Z$ increases (from 6 to 12) as the density increases due
to the (partial) crystallization of the packings, and we have $Z
\approx 6.00, 6.45$ and $7.96$ respectively for densities $\phi
\approx 0.64, 0.68$ and $0.72$ (rattlers are excluded when
computing $Z$).

Several representative packings obtained via the SLP
algorithm are also mapped onto the $\phi$-$\mathfrak{T}$
plane (red squares in Fig.~\ref{fig3}). \cite{T_independent} It
can be seen that jamming at $\phi<0.64$ is necessarily associated
with an increase in the degree of order, which is also indicated
by the increase of average contact number per particle (i.e., $Z
\approx 6.37$ for $\phi \approx 0.6$) and the decrease of the
fraction of rattlers (i.e., $\sim 1.1\%$ for $\phi \approx 0.6$).
The pair correlation function $g_2$ of a representative packing
configuration (with $N = 2235$) is shown in Fig.~\ref{fig4}.

The ``tunneled crystal'' packings are mapped onto the
$\phi-\mathfrak{T}$ plane (green up-triangles). The FCC tunneled
crystal possesses the highest translational order metric $\mathfrak{T}$,
the "disordered" Barlow tunneled crystal (a random stacking of layers of honeycomb-lattice sphere packings) possesses
the lowest $\mathfrak{T}$, and the ``zig-zag'' tunneled crystal (analog of the hexagonal-close
sphere packing) \cite{SalTunnel} is in between. The dashed blue lines show the spectrum of
packings generated by randomly filling the vacancies in the corresponding
``tunneled crystal'' packings, leading to perfect FCC, HCP and "disordered" Barlow packings at
the maximal density. \cite{SalReview} Several representative layered packings
are also mapped onto the $\phi$-$\mathfrak{T}$ plane (purple diamonds) in Fig.~\ref{fig3}.
The simple-cubic ($\phi = 0.5235\ldots$, blue left-triangle) and
body-centered-cubic ($\phi = 0.6801\ldots$, orange right-triangle) packings
are also shown in Fig.~\ref{fig3}. Note the SC and BCC packings are not collectively jammed.

\section{Conclusions and Discussion}

In summary, we have shown that jammed sphere packings can be produced
with a narrow distribution centered at any $\phi$ over a wide
range $\phi \in [0.6,~ 0.74048\ldots]$ with a diversity of disorder,
by exploring the algorithmic variability of packing protocols.
This suggests that any temptation to select a specific $\phi$ value as uniquely
significant or universal (e.g., $\phi \approx 0.64$) {\it based on its
frequence of occurrence} is primarily due to an
inadequate sampling of the full range of algorithmic
richness and variability of packing protocols.
Our packings are characterized by a ``geometric-structure'' approach,
i.e., they are mapped onto the $\phi$-$\mathfrak{T}$ plane; and
we have shown that moving away from the MRJ density in both
directions (i.e., lower or higher $\phi$) leads to a higher degree
of order and a larger average contact number.

It is claimed that the protocol (e.g., the athermal relaxation
method used in Ref.~\onlinecite{OHern}) in which the jammed states
of soft spheres are weighted by the volume of their basins of
attraction has a clear interpretation in terms of the energy
landscape, e.g., an ensemble of jammed states is considered. In
fact, the conceptual framework of such a protocol (e.g., the
energy landscape picture and the ensemble of inherent structures)
was introduced by one of us over 40 years ago. \cite{St64, St82,
St85} Although this approach is well-defined and theoretically
possible, it is inevitable that all protocols sample the energy
landscape in their own biased fashion. Therefore, there is no
compelling mathematical criterion for selecting one protocol over
the others. For example, from completely random initial
configurations (i.e., Poisson distribution of points), it is found
the jamming density sharply peaks at $\phi \approx 0.64$.
\cite{OHern} However, it was shown nearly a decade ago that a
much wider density range can be achieved for both monodisperse and
polydisperse sphere packings using the Lubachevsky-Stillinger
packing algorithm; \cite{SalMRJ, Truskett00, SalMRJ2, Ka02} and
recently similar results for polydisperse spheres have been
obtained \cite{Cha10} using the same athermal protocol as in
Ref.~\onlinecite{OHern}. More recently, we have devised a novel
(athermal) linear-programming packing protocol that enables one to
obtain a spectrum of inherent structures ranging from disordered jammed
packings up to the maximal density packings starting from
\textit{random initial configurations}. \cite{To10} All of these
results suggest that there is no ``universal'' protocol that can
generate all possible jammed states in an unbiased fashion.

Moreover, it is not clear how the jammed states should be
weighted in an ensemble. It has been suggested that the volume of
the basins of attraction for systems starting from Poisson
distributions of interacting points \cite{OHern} should be used
for the weighting. However, in reality virtually no jammed systems
are experimentally produced using a Poisson distribution as an initial
configuration. In addition, the jammed packings produced either
numerically or experimentally always contain a small but different
fraction of rattlers. Strictly speaking, these jammed states are
not single points on the energy landscape but bounded regions
with dimensions equal to the number of degrees of freedom of the
rattlers. The dimension of the jamming basins are different, and
therefore it is ambiguous to consider their volumes for weighting
and to characterize them on the same footing. Ideally, each
jamming basin should be characterized individually as emphasized
in the ``geometric-structure'' approach. Removing the rattlers
will not affect the jamming nature of the packings, but leads to
the lower jamming density. This further adds ambiguity to a
density-alone characterization of jammed packings, which is
employed in the ``ensemble'' approach.

We also would like to note that as exercised the ``ensemble''
approach has focused on protocols that disproportionately generate
disordered packings, presumably because such packings have a high
frequency of occurrence in typical experiments or simulations. The
specific protocols employed discriminate against the ordered structures
such as the FCC packing (and its stacking variants) and the
tunneled crystals \cite{SalTunnel} and result in an inappropriate
narrow vision of the set of all possible collectively jammed
packings. Thus, the physical relevance of jammed packings should
not be determined based on their frequency of occurrence in
experiments or simulations, which again are protocol dependent. 

From our point of view, the ``ensemble'' and ``geometric-structure''
approaches do not conflict with each other but rather are
complementary. For example, the ``geometric-structure'' approach
characterizes individual packing configurations drawn from
well-defined ensembles. However, in our experience, the
``geometric-structure'' approach can always provide nontrivial
solutions when the ``ensemble'' approach breaks down, such as the
possible identification of disordered jammed two-dimensional
packings discussed in the Introduction. Therefore, we believe that
the ``geometric-structure'' approach, when utilized in the broad
context of a full set of available protocols, has a capability to
discover, integrate and characterize packing structures that go
well beyond the disordered set. We emphasize that all of the
aforementioned issues have been addressed in the literature.
\cite{SalMRJ, Truskett00, SalMRJ2, SalReview, SalComment} In this
paper, we have examined the applications of the
``geometric-structure'' approach in a broader context by
amplifying and quantifying some of these issues.


An interesting open question that naturally arises is whether $\phi \approx 0.60$
is the lower limit on the density of jammed amorphous sphere packings?
We note that the distribution of density for $N\sim2500$ shown in Fig.~\ref{fig1}(a) can
be fitted with a Gaussian with mean $\phi = 0.602$ and standard
deviation $\sigma = 0.002$. This strongly indicates that there is a small but finite probability
of finding jammed packings with even lower densities. Given
enough number of trials, such packings could be obtained in principle.
It is noteworthy that the density $\phi = \frac{\sqrt2
\pi}{9} = 0.49365\ldots$ associated with the ``tunneled crystals''
is likely to be the threshold (i.e., lowest possible) density for
strictly jammed sphere packings in three dimensions. \cite{SalTunnel} However,
neither the LS nor the SLP algorithms as normally implemented is
able to produce such packings, presumably because both algorithms
tend to densify the packing and jamming is a consequence of their
``compression'' nature. Lower $\phi$ is attainable via the SLP
algorithm because jamming is achieved through local
reconfigurations. However, it is currently not designed to find
highly unsaturated jammed packings, such as the tunneled crystals.
Therefore, we see no reason that jammed amorphous sphere packings
with even lower density cannot be produced via carefully designed
protocols.

A limitation of current protocols designed to produce jammed packings 
is that they inevitably lead to packings with high densities. It is highly
desirable to devise protocols that explicitly take into account
the requirement of jamming as well as other packing
characteristics (e.g., $\phi$ and $Z$). One possible approach is
to delineate the conditions for jamming (and other
characteristics) in a quantitative way and to include them as
constraints of an optimization problem. We have
suggested possible solutions to the problem of producing
low-density jammed amorphous packings in Ref.~\onlinecite{To10}.
In future research, we will focus on the development of such
packing protocols, which is a highly nontrivial challenge.


\begin{acknowledgments}
This work was supported by the National Science Foundation under
Grants DMS-0804431 and DMR-0820341.
\end{acknowledgments}

\clearpage

\newpage

\begin{figure}
\centering \caption{Histograms of jammed sphere packings that are
centered around different mean densities. Panel (a) shows packings
generated using the SLP algorithm with $\phi \approx 0.6$. Panels
(b), (c) and (d) show packings generated using the LS algorithm
with $\phi \approx 0.64$, $0.68$ and $0.72$, respectively. The
distributions become narrower as the system size increases.}
\label{fig1}
\end{figure}

\begin{figure}
\centering \caption{The standard deviation $\sigma$ of the density
distribution with mean $\phi \approx 0.66$ as a function of system
size $N$. System sizes $N=500, 1000, 2500, 5000$ and $10000$ are
used here. The linear fit for $\ln \sigma$ vs. $\ln N$ gives a
slope $k\approx -0.533$, which indicates that $\sigma$
approximately scales as $N^{-1/2}$ for large N. Such a
$\sigma$-$N$ relation is also observed for density distributions
with the other mean values studies here.} \label{fig2}
\end{figure}

\begin{figure}
\centering \caption{Order map for jammed sphere packings with
$N\sim2500$: translational order metric $\mathfrak{T}$ versus
packing density $\phi$. The dashed blue lines (which are
consistent with the qualitative trends indicated in the order maps
discussed in Ref. \onlinecite{SalReview}) show the spectrum of
packings generated by randomly filling the vacancies in the
corresponding tunneled crystal pckings, which leads to perfect
FCC, HCP and Barlow packings at the maximal density. The
simple-cubic (SC) and body-centered-cubic (BCC) packings (not
jammed) are also shown.} \label{fig3}
\end{figure}

\begin{figure}
\centering \caption{Pair correlation function $g_2$ and average
contact number $Z$ of representative jammed sphere packings at
different densities, where $D$ is the sphere diameter.}
\label{fig4}
\end{figure}

\clearpage
\newpage

\setcounter{figure}{0}

\begin{figure}
$\begin{array}{c@{\hspace{1.5cm}}c}
\includegraphics[height=4.5cm, keepaspectratio]{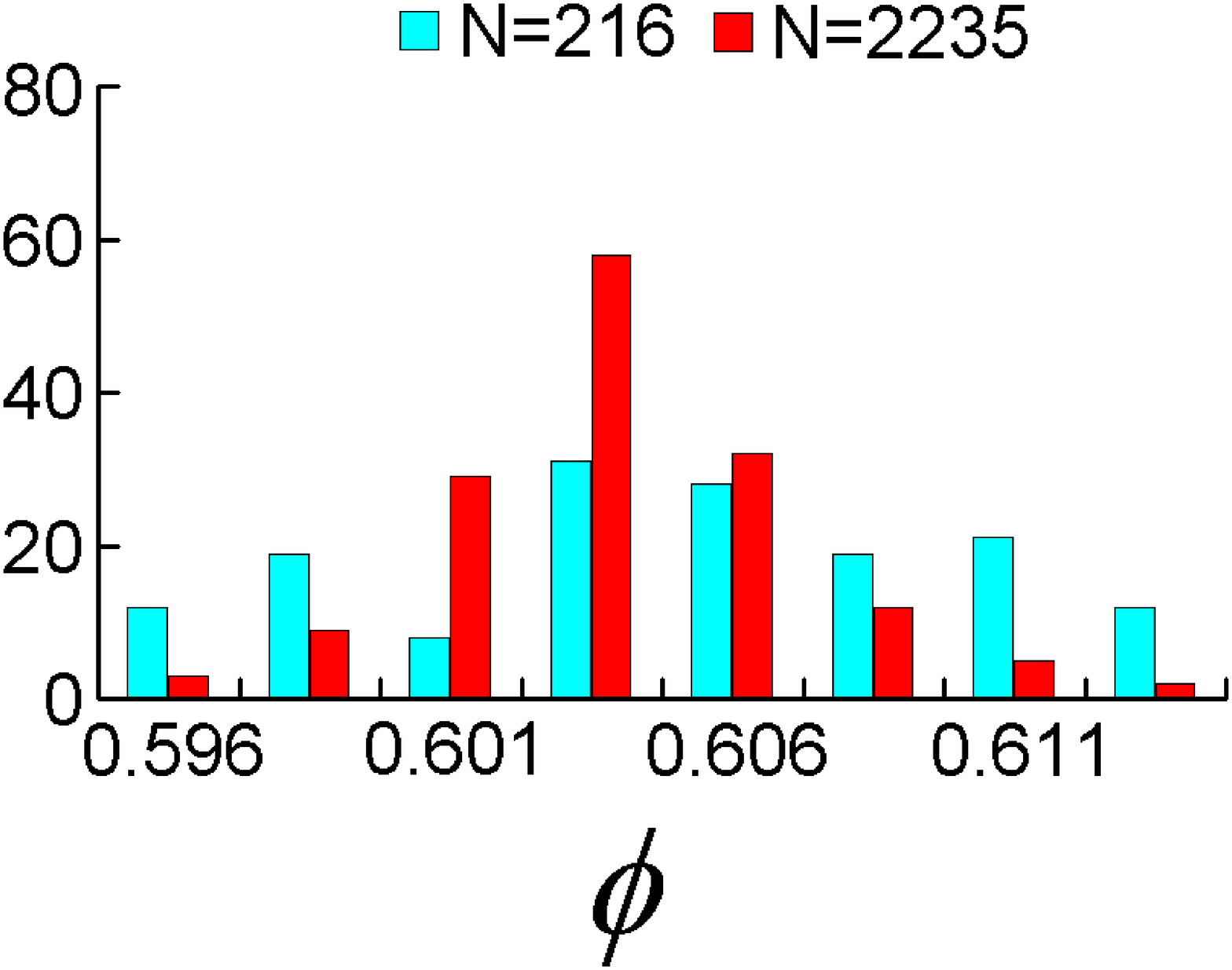} &
\includegraphics[height=4.5cm, keepaspectratio]{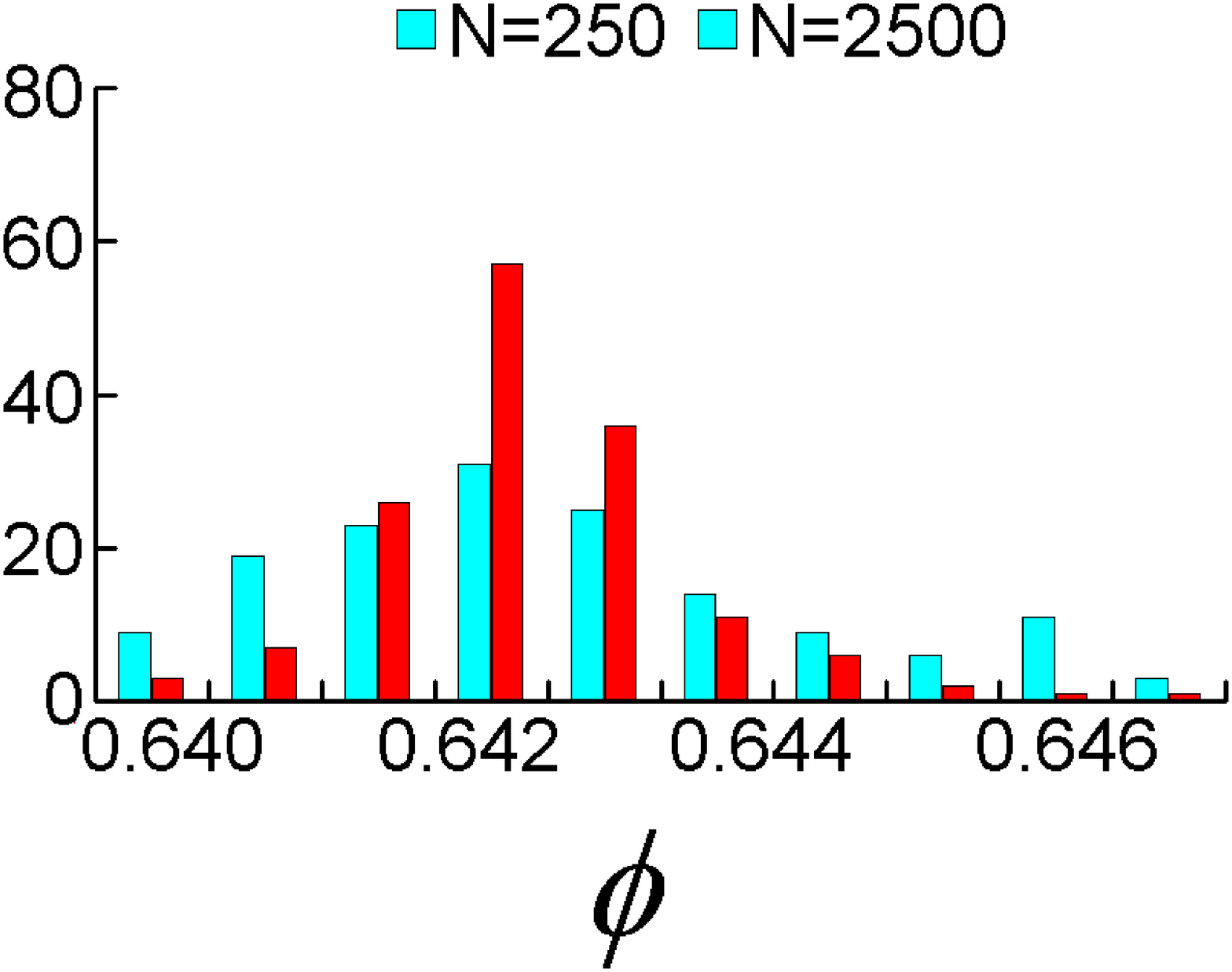} \\
\mbox{(a)} & \mbox{(b)} \\\\
\includegraphics[height=4.5cm, keepaspectratio]{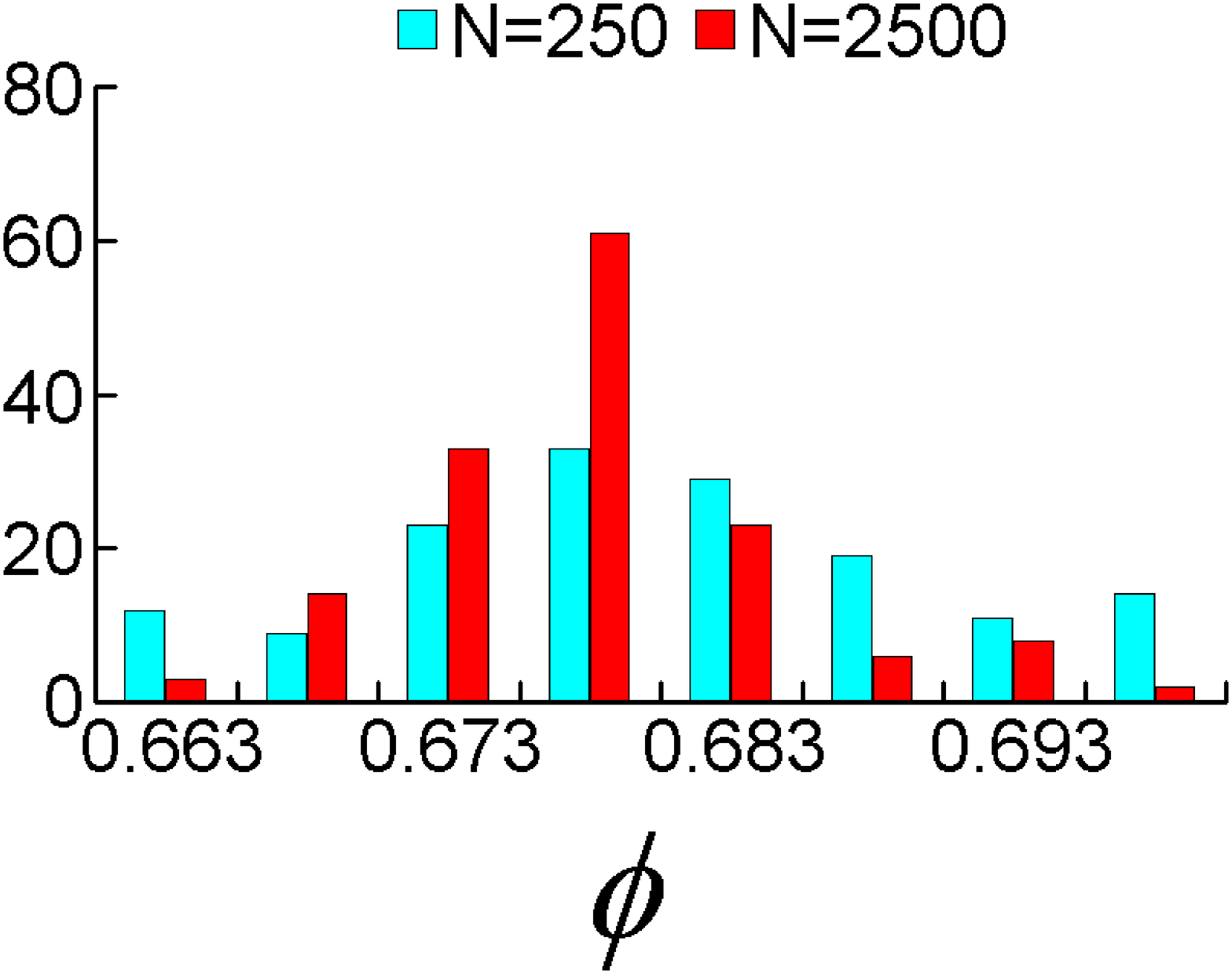} &
\includegraphics[height=4.5cm, keepaspectratio]{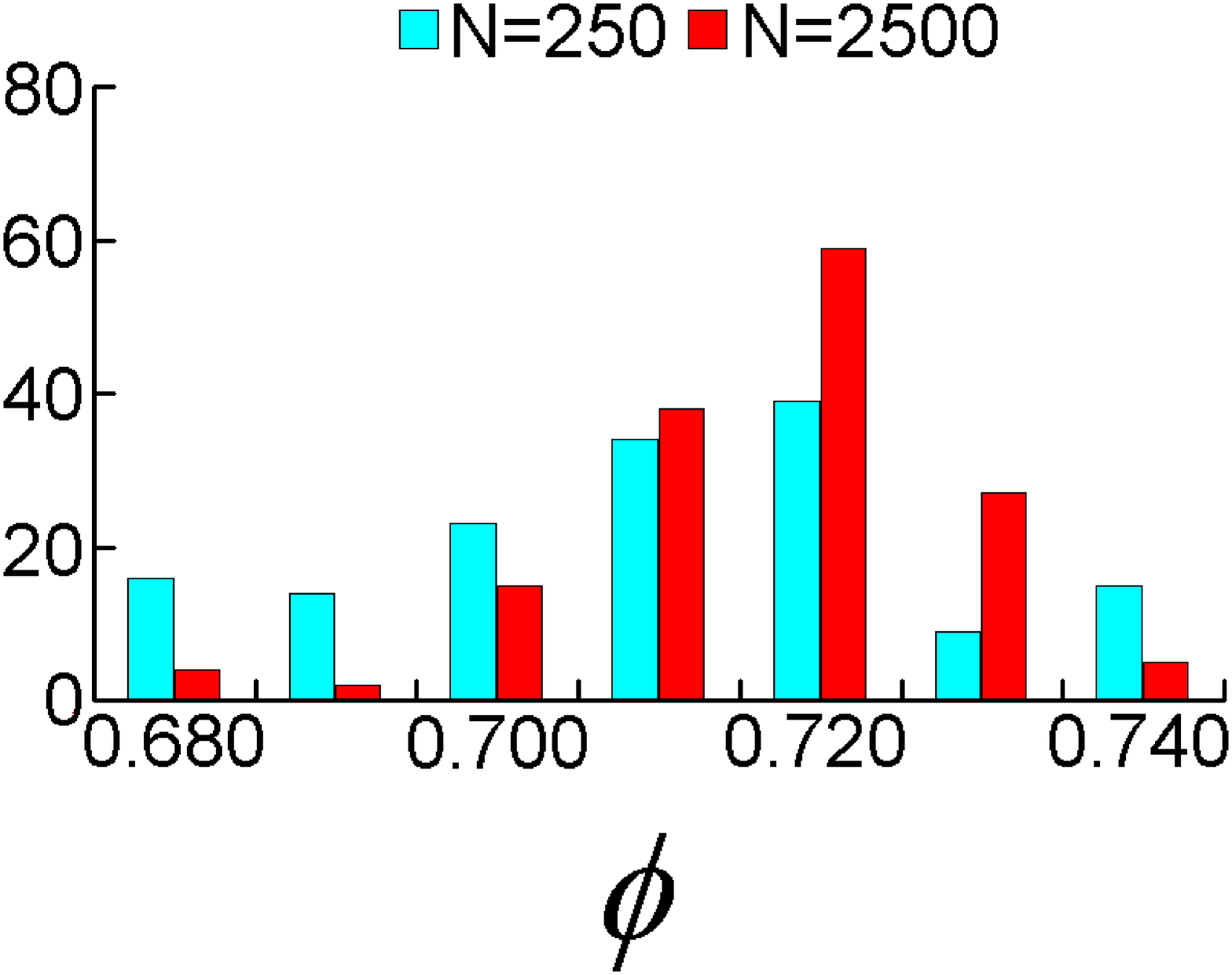} \\
\mbox{(c)} & \mbox{(d)}
\end{array}$
\centering \caption{Jiao, Stillinger, Torquato}
\end{figure}

\clearpage
\newpage

\begin{figure}
\includegraphics[width = \textwidth, clip=true]{fig2.eps}
\centering \caption{Jiao, Stillinger, Torquato}
\end{figure}

\clearpage
\newpage

\begin{figure}
\includegraphics[width = \textwidth, clip=true]{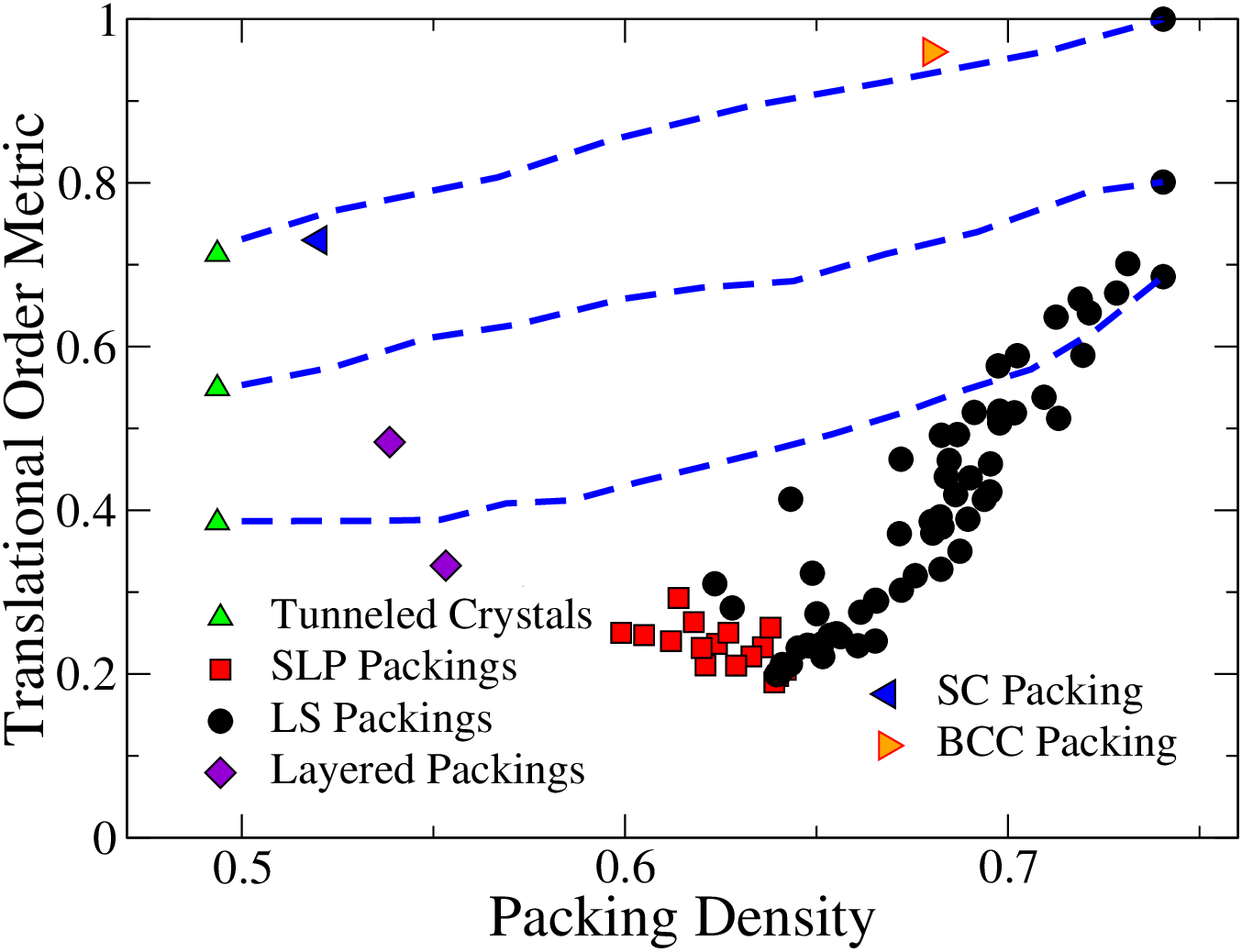}
\centering \caption{Jiao, Stillinger, Torquato}
\end{figure}

\clearpage
\newpage

\begin{figure}
\includegraphics[width = \textwidth, clip=true]{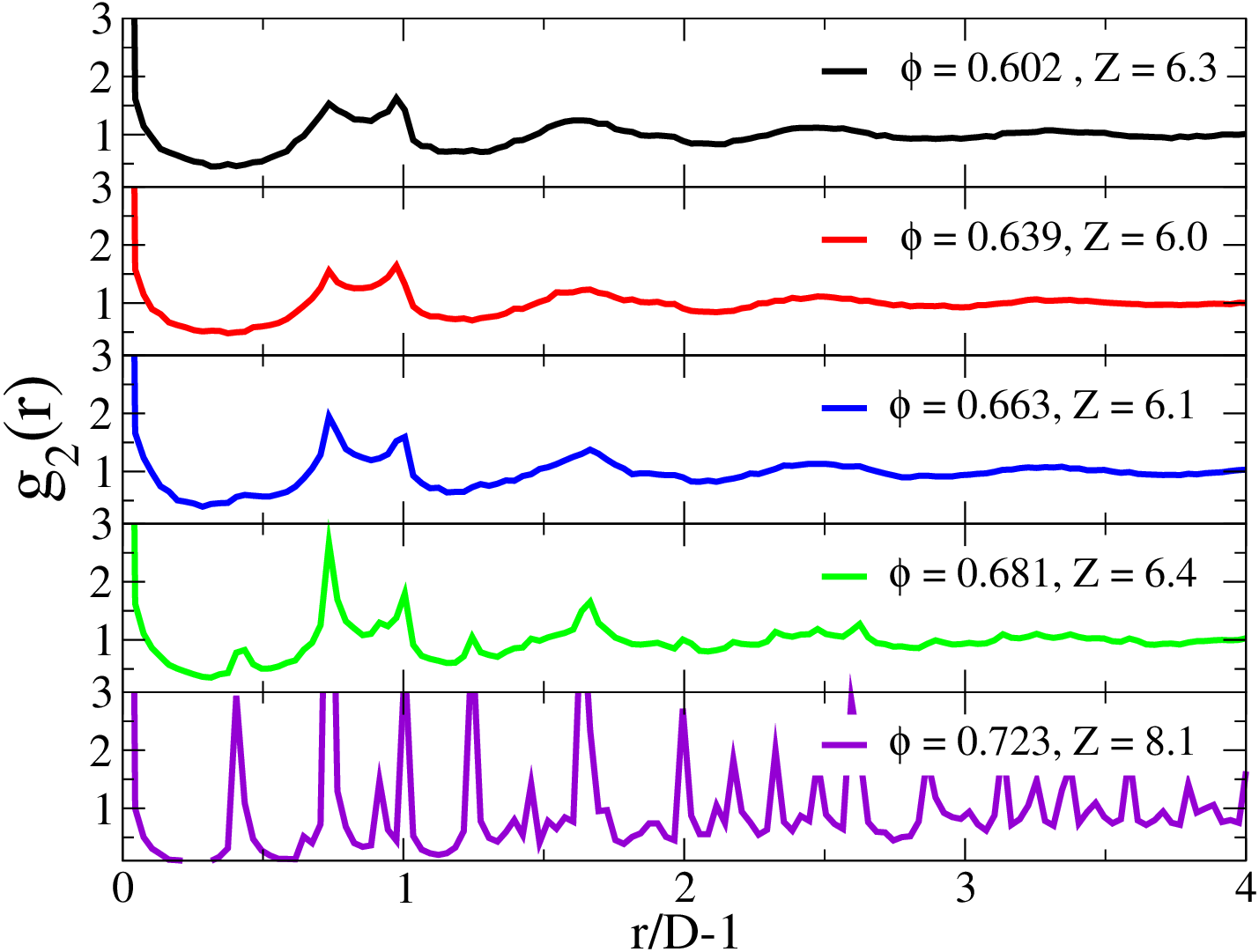}
\centering \caption{Jiao, Stillinger, Torquato}
\end{figure}


\begin{thebibliography}{10}
\bibitem{T_independent}
Note that we have also computed the ``crystal-independent''
translational order metric $T^*$ defined in Ref.
\onlinecite{Truskett00} for typical packings for $0.6 \le \phi \le
0.64$. As expected, these results for $T^*$ are positively
correlated with those for the aforementioned ``crystal-dependent''
translational order metric $\mathfrak{T}$ and hence these two
translational order metrics are consistent with one another.
\end{thebibliography}
\end{document}